\documentclass[12pt,preprint]{aastex}
\usepackage{enumerate}
\usepackage{float}
\usepackage{placeins}
\usepackage{amsmath}
\usepackage{natbib}
\usepackage{epstopdf}
\usepackage{lscape}
\usepackage{verbatim}
\newcommand{\HI}{H\,{\sc i}~}
\newcommand{\HII}{H\,{\sc ii}~}

\begin{document}

\title{Nebular metallicities in two isolated Local Void dwarf galaxies}
\shorttitle{Nebular metallicity in Local Void dwarf galxies}
\shortauthors{Nicholls et~al.}

\author{ David C. Nicholls\altaffilmark{1}, Helmut Jerjen\altaffilmark{1}, Michael A. Dopita\altaffilmark{1}$^,$\altaffilmark{2}, \& Hassan Basurah\altaffilmark{2},}
\email{david@mso.anu.edu.au}
\altaffiltext{1}{Research School of Astronomy and Astrophysics, Australian National University, Cotter Rd., Weston ACT 2611, Australia }
\altaffiltext{2}{Astronomy Department, King Abdulaziz University, P.O. Box 80203, Jeddah, Saudi Arabia}

\begin{abstract}
Isolated dwarf galaxies, especially those situated in voids, may provide insight into primordial conditions in the Universe and the physical processes that govern star formation in undisturbed stellar systems. The metallicity of \HII regions in such galaxies is key to investigating this possibility. From the SIGRID sample of isolated dwarf galaxies, we have identified two exceptionally isolated objects, the Local Void galaxy [KK98]246 (ESO 461-G036) and another somewhat larger dwarf irregular on the edge of the Local Void, MCG-01-41-006 (HIPASS J1609-04).  We report our measurements of the nebular metallicities in these objects. The first object has a single low luminosity \HII region, while the second is in a more vigorous star forming phase with several bright \HII regions. We find that the metallicities in both galaxies  are typical for galaxies of this size, and do not indicate the presence of any primordial gas, despite (for [KK98]246) the known surrounding large reservoir of neutral Hydrogen.
\end{abstract}

\keywords{galaxies: individual: [KK98]246 (ESO 461-G036); MCG-01-41-006 (HIPASS J1609-04) --- galaxies: dwarf --- galaxies: irregular --- galaxies: evolution --- galaxies: formation --- H~\textsc{ii} regions --- ISM: abundances}

\section{Introduction}

The physical isolation of a dwarf galaxy is often argued to be a major contributor in preserving evidence of intergalactic \HI gas in its primordial condition.  Isolated dwarf galaxies are thought to be very slow to enrich their neutral gas clouds, compared to the much more rapid evolution of larger galaxies \citep[e.g.,][]{Mateo98}.  Their relatively shallow potential wells are argued to be inefficient in retaining the heavy element abundances generated by supernovae \citep[e.g.,][]{Kunth00}.  This picture is consistent with the mass-metallicity relation \citep[e.g.,][]{Lee03, Tremonti04, Lee06}, which shows that smaller galaxies generally exhibit lower nebular metallicity than larger ones.  Investigating the nature of this relation was one of the motivations behind the SIGRID sample of isolated gas-rich dwarf galaxies \citep{Nicholls11}, which, \emph{inter alia},  seeks to clarify the behaviour of the mass-metallicity relation at low galactic masses. It has also been suggested that dwarf galaxies residing in low density environments (e.g., in voids) have lower metallicities, and in some cases, very low mass-to-light ratios, compared to similar objects in higher density regions \citep{Pustilnik07,Pustilnik11a,Pustilnik11b}. Therefore is is useful to investigate whether spatially isolated galaxies in voids are in any way unusual, in terms of mass-to-light ratio or metallicity, compared to similar objects in denser environments. 

We have identified two isolated gas-rich dwarf galaxies in the SIGRID sample located within the Local Void for investigation, [KK98]246 (ESO 461-G036), hereafter ``KK246'' and MCG-01-41-006 (HIPASS J1609-04), hereafter ``J1609''. KK246 is relatively nearby \citep[6.4 Mpc, ][]{Tully08}. J1609-04 is somewhat more massive and luminous, but is over twice as distant, at 14.82 Mpc \citep{Nicholls11}. We present the spectra and nebular metallicity results for the two galaxies, and compare them with data from other investigations of void galaxies.

This paper is organised as follows. In sections (2) and (3) we provide a summary of what is known about the two galaxies.  In section (4) we detail our observations. In section (5) we present the flux-calibrated spectra from 3600\AA \ to 7000\AA \ at resolutions of 3000 (blue) and 7000 (red). In section (6) we present the methods and results for our metallicity analyses; we list the observed line fluxes; and nebular metallicities and ionization parameters using the revised strong line methods described in \cite{Dopita13}. In Section (7) we present comparisons with other surveys of similar galaxies for mass-to-light ratios and mass-metallicities. In section (8) we discuss the implications for dwarf galaxy evolution of these observations. Section (9) presents our conclusions.

\section{KK246}

\subsection{Description}
KK246 is a particularly interesting object in the SIGRID sample  as it has been identified as the most isolated dwarf galaxy known in the Local Volume (D$\lesssim$10\,Mpc, Karachentsev et al. 2013). It is situated a few Mpc within the Local Void \citep{Tully08, Kreckel11, Nasonova11}.  This void is large, at least 23Mpc in radius \citep{Tully08}, and KK246 has no apparent neighbours within $\sim 3$\,Mpc \citep{Karachentsev04}. Assuming a spatial peculiar velocity of 100-200\,km\,s$^{-1}$ this translates into a time of 15-30 Gyr since the last possible galaxy-galaxy encounter. As such, the object is an ideal laboratory in which to study galaxy evolution and self-enrichment in isolation. It is of particular interest in understanding the intrinsic evolutionary processes in the formation of this small galaxy, as almost certainly it has never been affected by tidal disruption by nearby large galaxies, nor by enrichment of the intergalactic medium (IGM) from which it formed by outflows from such galaxies. It has no detectable adjacent dwarf galaxies, unlike the low metallicity LSB galaxies in the Lynx-Cancer void at 18\,Mpc investigated by \cite{Pustilnik11b}.

Previous observations of KK246 have shown that the stellar component is embedded in a large surrounding \HI cloud (Kreckel et al. 2011) 
and contains an old stellar population \citep{Karachentsev06}.  Although it currently has only one small, low-luminosity star-forming \HII region, we have been able to measure the nebular metallicity of that region using the WiFeS IFU spectrograph on the ANU 2.3m telescope at Siding Spring Observatory. This paper reports the first measurement of the nebular spectrum and resultant nebular metallicities. 

\subsection{Location}

Using HST imagery and photometry, \cite{Karachentsev06} measured the TRGB distance to KK246 as 7.83 Mpc. \citet{Tully08} use a more recent calibration of the T$_{rgb}$ scale from \cite{Rizzi07} and give a distance of 6.4 Mpc.  With either distance value, KK246 is within the Void \citep{Tully08}. In the calculations in this paper, we have used the value from Tully et~al., but our conclusions do not depend on which value is adopted.

The object's isolation is indicated by the low value of the Tidal Index $\Theta=-2.2$ \citep{Karachentsev04} and the SIGRID $\Delta$ index \citep[not measurable: no potential tidally effective objects within 10\degr \ or  $\sim$1.3\,Mpc radius, ][]{Nicholls11}, implying no tidal interference from any nearest neighbour galaxy within a Hubble time\footnote[1]{$\Theta$ takes into account galaxies whose mass  is known, at large angular distances from the tidally influenced object, whereas $\Delta$ is sensitive to dwarf galaxies within 10\degr \ in the sky, whose mass  has not been directly measured, but whose flow-corrected redshift and luminosity are known.}. \citet{Karachentsev04} have concluded that there are no potential interacting galaxies within 3 Mpc.  KK246 is also listed in the ``Local Orphan Galaxies'' catalog \citep{Karachentsev11} as having the highest recorded isolation index (log$_{10}$ (ii) = 3.22) using an alternative isolation measure.

\subsection{Extended \HI region}

KK246 is one of over 4500 \HI objects detected in the HIPASS survey  \citep[HIPASS J2003-31, ][]{Meyer04} and was identified with the optical counterpart galaxy ESO 461-G036  \citep{Doyle05}.  It was subsequently observed using the VLA at 21cm at higher resolution by \citet{Kreckel11}, who found that KK246 is surrounded by a very extended \HI region, at least ten times the diameter of the stellar disk. 

\citet{Kreckel11} find a large dynamical mass ($4.1\times10^9$ M$_\odot$) and consequently one of the highest measured dynamic-mass-to-light ratios, $M_{\rm dyn}/{L}_{B}=89$, further emphasising the unusual nature of this galaxy.  Other noteworthy high 
dynamic-mass-to-light ratio objects include NGC 2915 \citep[$M_{\rm dyn}/{L}_{B}=76$,][]{Meurer96}, ESO 215-G?009 \citep[$M_{\rm dyn}/{L}_{B}=22$, ][]{Warren04}, and NGC 3741 \citep[$M_{\rm dyn}/{L}_{B}=107$, ][]{Begum05}. However these objects have \HII regions (using GALEX FUV flux as a diagnostic) that are either much brighter (NGC 2915, 3741) or more spatially extended (ESO 215-G?009), than KK246. 

\subsection{Stellar population}
\begin{figure}[h]
\includegraphics[width=0.5\hsize]{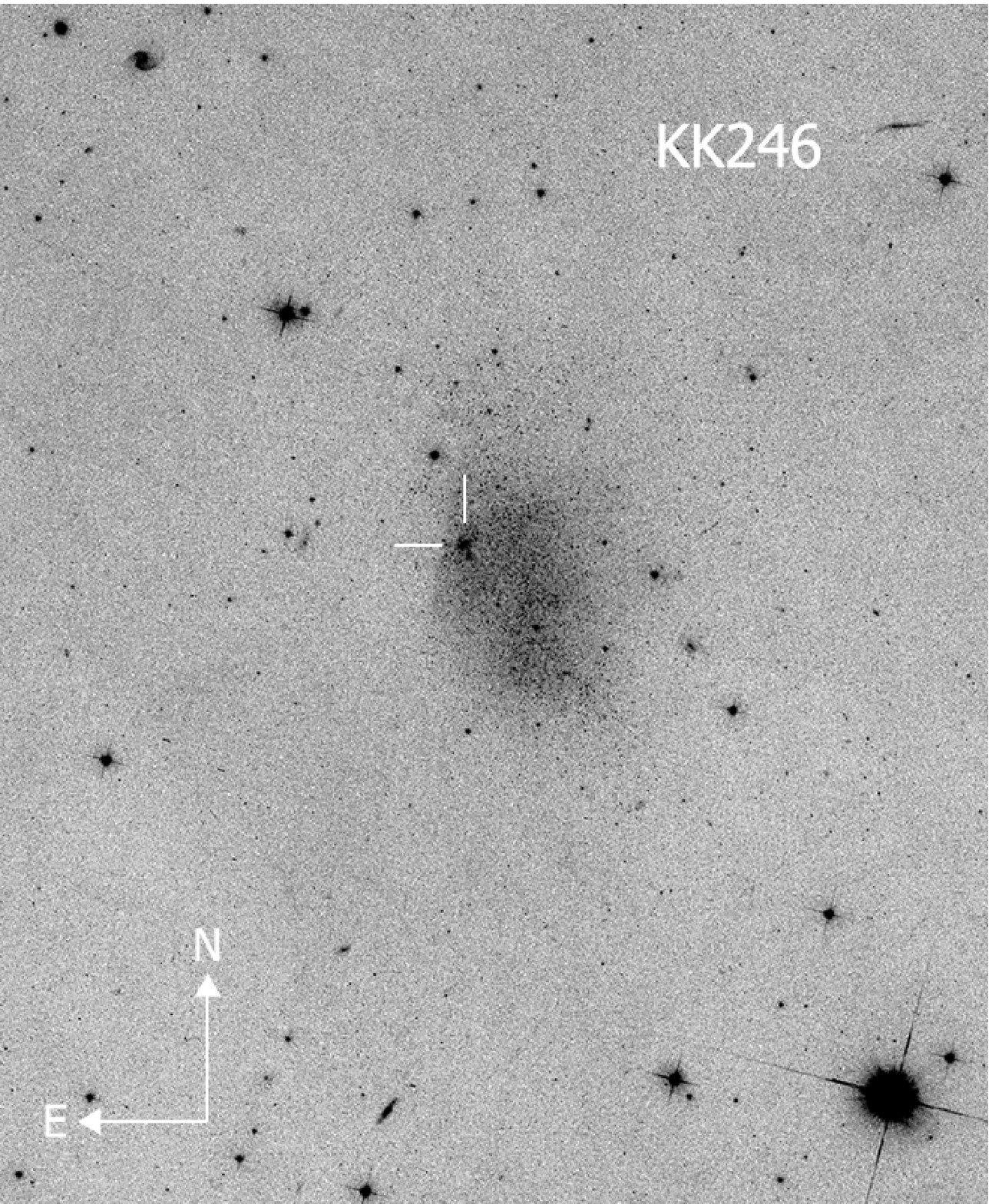}
\includegraphics[width=0.5\hsize]{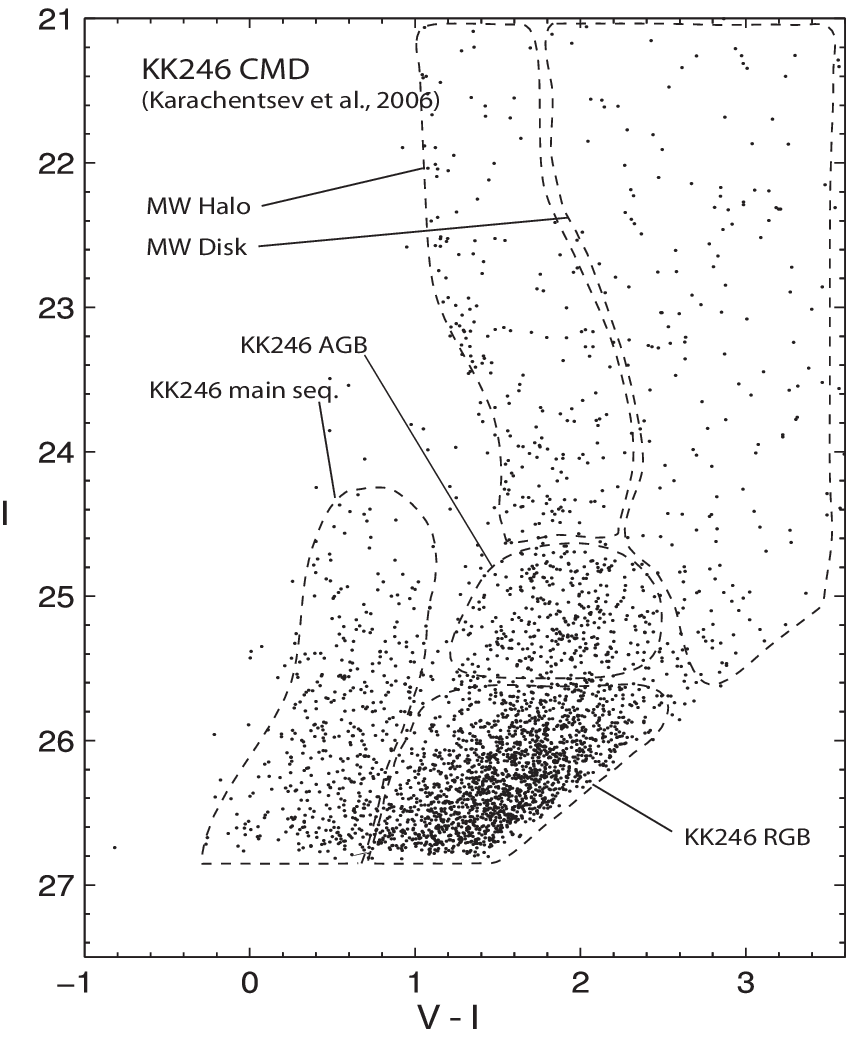}
\caption{Left Panel: HST V-band image of KK246. The only \HII region in the galaxy is marked  to the NE of the centre. Right Panel: HST (V-I,I) color-magnitude diagram derived from \cite{Karachentsev06}, showing evidence of at least two stellar populations. The transition from the RGB to the AGB stars occurs around I $\sim$ 25.6 mag.}\label{fig_1}
\end{figure}

As part of a survey of Local Volume galaxies aimed at measuring TRGB distances, \citet{Karachentsev06} observed KK246 using the Hubble Space Telescope. The left panel of Figure \ref{fig_1} shows the composite F606W-band image, obtained from the HST archive, from those observations.  The location of the \HII region observed in this work is marked.

\FloatBarrier 

\cite{Karachentsev06} resolved the brightest two magnitudes of the color-magnitude diagram for KK246 from HST WFC3 images, which shows clear evidence of an old stellar population.  The large distance to KK246  makes star formation modelling difficult, which rules out calculating the star formation history in detail, but by comparing its CMD diagram with the CMD data for 60 nearby dwarf galaxies in the ANGST survey from \cite{Dalcanton09}, and their star formation histories analysed by \cite{Weisz11}, it appears likely that KK246 has had at least three epochs of star formation, including the present one involving the \HII region.  The right panel of Figure \ref{fig_1} shows our interpretation of the CMD from \cite{Karachentsev06}.

\subsection{UV, visible, H$\alpha$ and H-band imagery}

KK246 is detected in both the NUV and FUV images from GALEX (see Table 1).  The low FUV flux [luminosity] of 28.2 $\mu$Jy implies that there is only a small population of hot young stars currently present in the galaxy. Using equations 17 and 18 from \cite{Karachentsev13} we infer a star formation rate of log(SFR)=-2.43.  This is consistent with our H$_\alpha$ observations (see below).  Figure \ref{fig_2} (bottom, left) shows the composite GALEX image. Although the resolution and sensitivity are marginal, it is apparent that the brightest UV object coincides with the \HII region.

The same figure also shows the visible color image from DSS2 (top left), the single \HII region from a H$\alpha$ + R-band WiFeS data cube, (top right), and an H-band image from \cite{Kirby08} (bottom right).  The small \HII region corresponds to the point marked in the HST image (Figure \ref{fig_1}) above.  The physical extent, spectrum (see below) and H$\alpha$ flux confirm that this is an \HII region and not a planetary nebula, unlike the object discussed by \cite{Makarov12} in the isolated dwarf spheroidal galaxy KKR25.

\begin{figure}[htpb]
\includegraphics[width=1.0\hsize]{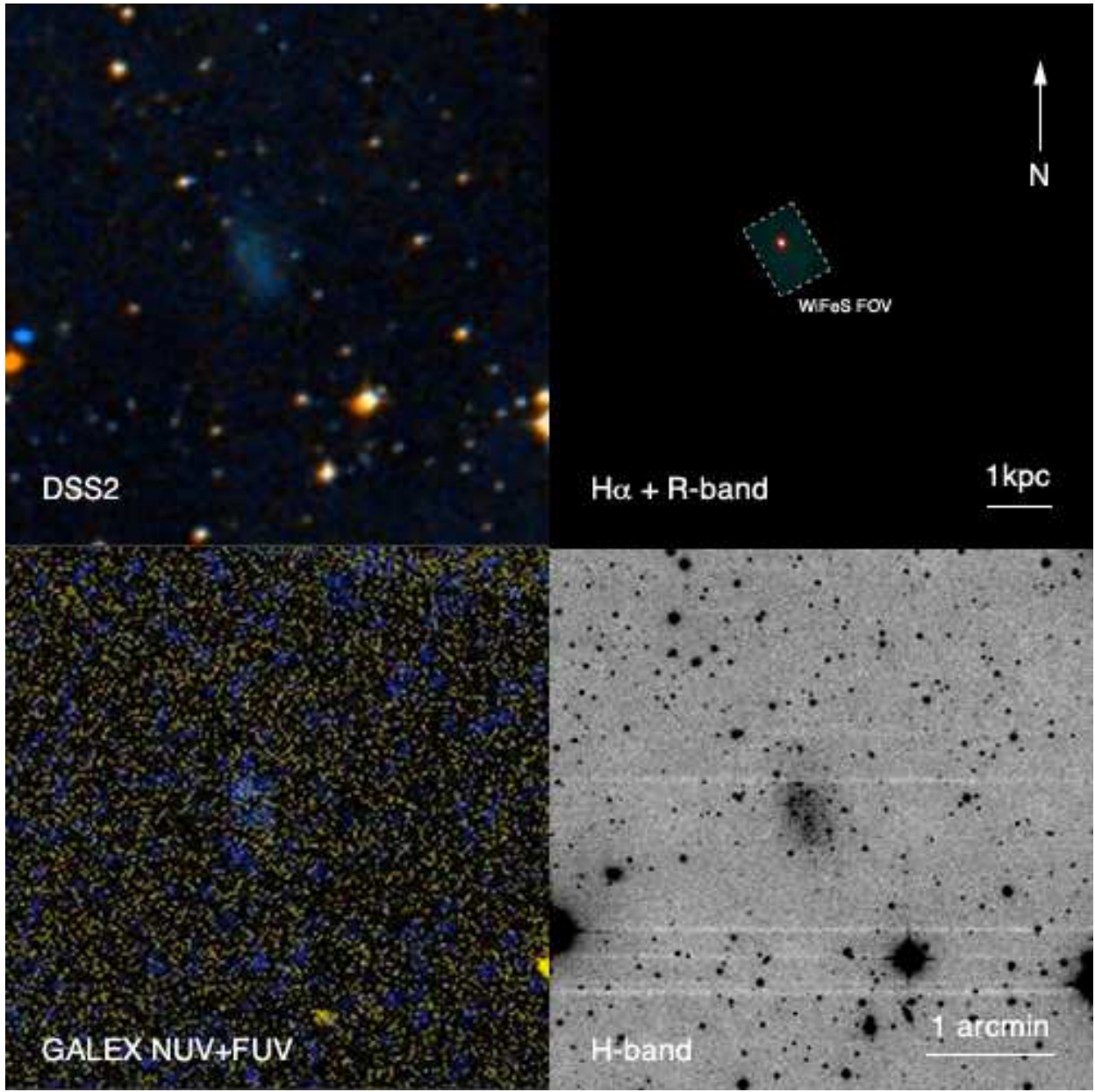}
\caption{KK246 in visible, H$\alpha$/R-band, H-band, and GALEX NUV + FUV , all to the same scale}\label{fig_2}
\end{figure}

Table 1 summarises information available on KK246.

\begin{deluxetable}{lrl}
\centering
\tabletypesize{\footnotesize}
\tablecaption{Compendium of UV, Optical, Near-IR, and HI Parameters for KK246}
\tablehead{
\colhead{Parameter} & \colhead{Value} & \colhead{Reference} \\
}
\startdata
R.A. (J2000)					&	20 03 57.4	& \cite{Lauberts82}\\
Dec (J2000)					&	--31 40 54		& \cite{Lauberts82}\\
l (deg)						&	9.72864 	&  derived from \cite{Lauberts82}\\
b (deg)						&	--28.36894 &  derived from \cite{Lauberts82}\\
\HII region R.A. (J2000)			&	20 03 57.7	& This work \\
\HII region Dec (J2000)			&	--31 40 48		& This work \\
Dist (Mpc)						& 	6.4$^a$		& \cite{Tully08} \\
m$_B$ (mag)					& 	17.22		& \cite{Tully08}\\
$A_B$ (mag)					& 	1.10			& \cite{Schlafly11} \\
E$_{B-V}$ (mag.)				& 	0.154		& \cite{Schlafly11} \\
M$_{B,0}$ (mag)				& 	--12.91		&  This work\\
log$_{10}$(L$_B$/L$_\sun$)		& 	7.524		 & This work\\
m$_H$ (mag)					&	13.9$\pm$0.2	& \cite{Kirby08} \\
M$_{H,0}$ (mag)				&	--15.9$\pm$0.3	& \cite{Kirby08} \\
FUV flux ($\mu$Jy)				&	 28.18$\pm$3.89  &  GALEX Data Release 6/7.\\ 
NUV flux ($\mu$Jy)				&	 26.05$\pm$3.72  &  GALEX Data Release 6/7\\
Total H~\textsc{i} flux (Jy km  s$^{-1}$)	&	7.3	& \cite{Kreckel11}\\
							&	7.5			& \cite{Meyer04}\\
W$_{20}$ (km  s$^{-1}$)			&	93 			& \cite{Kreckel11}\\
							&	104.5		& \cite{Meyer04} \\
W$_{50}$ (km  s$^{-1}$)			&	71.6		         & \cite{Meyer04} \\							
log$_{10}(\mathcal{M}_{HI}/{\mathcal{M}}_{\odot}$)&	8.02$\pm$0.03		& \cite{Kreckel11}\\
							&	8.03			& derived from \cite{Meyer04}\\
log$_{10}(\mathcal{M}_{\ast}/{\mathcal{M}}_{\odot}$)	&	7.7$\pm$0.2	&  \cite{Kirby08} \\
M$_{\rm H\textsc{i}}$/L$_B$		&	4.656		& This work\\
$\Theta$ tidal index				&      --2.2			& \cite{Karachentsev04}\\
$\Delta$ tidal index				&	(nil)		& \cite{Nicholls11}\\
log$_{10}$(ii) isolation index		&	3.32			& \cite{Karachentsev11}\\
\enddata
\tablecomments{
(a) \cite{Tully08} use a more recent calibration of the T$_{rgb}$ scale from \cite{Rizzi07} and give a distance of 6.4 Mpc, compared to the value of 7.83 Mpc from \cite{Karachentsev06}. This does not affect the conclusions reached here.}\\
\end{deluxetable}\label{table_1}
\FloatBarrier

\section{HIPASS J1609-04}
\subsection{Description and location}

Another interesting galaxy from the SIGRID sample in this context is HIPASS J1609-04. It is located on the edge of the Local Void, 7.65Mpc from its center  \citep{Nasonova11}. It is one of the more isolated galaxies in the SIGRID sample \citep{Nicholls11}, with a $\Delta$ tidal index of -2.9.  It has an isolation index of 2.38 \citep{Karachentsev11}, compared to 3.22 for KK246. Its distance of 14.82\,Mpc is estimated from local flow data for the Virgo, Shapley and GA fields. Its baryonic mass is about four times larger than KK246, and it is substantially more actively star forming.

\subsection{UV, visible, H$\alpha$ and H-band imagery}

\begin{figure}[htpb]
\centering
\includegraphics[width=0.85\hsize]{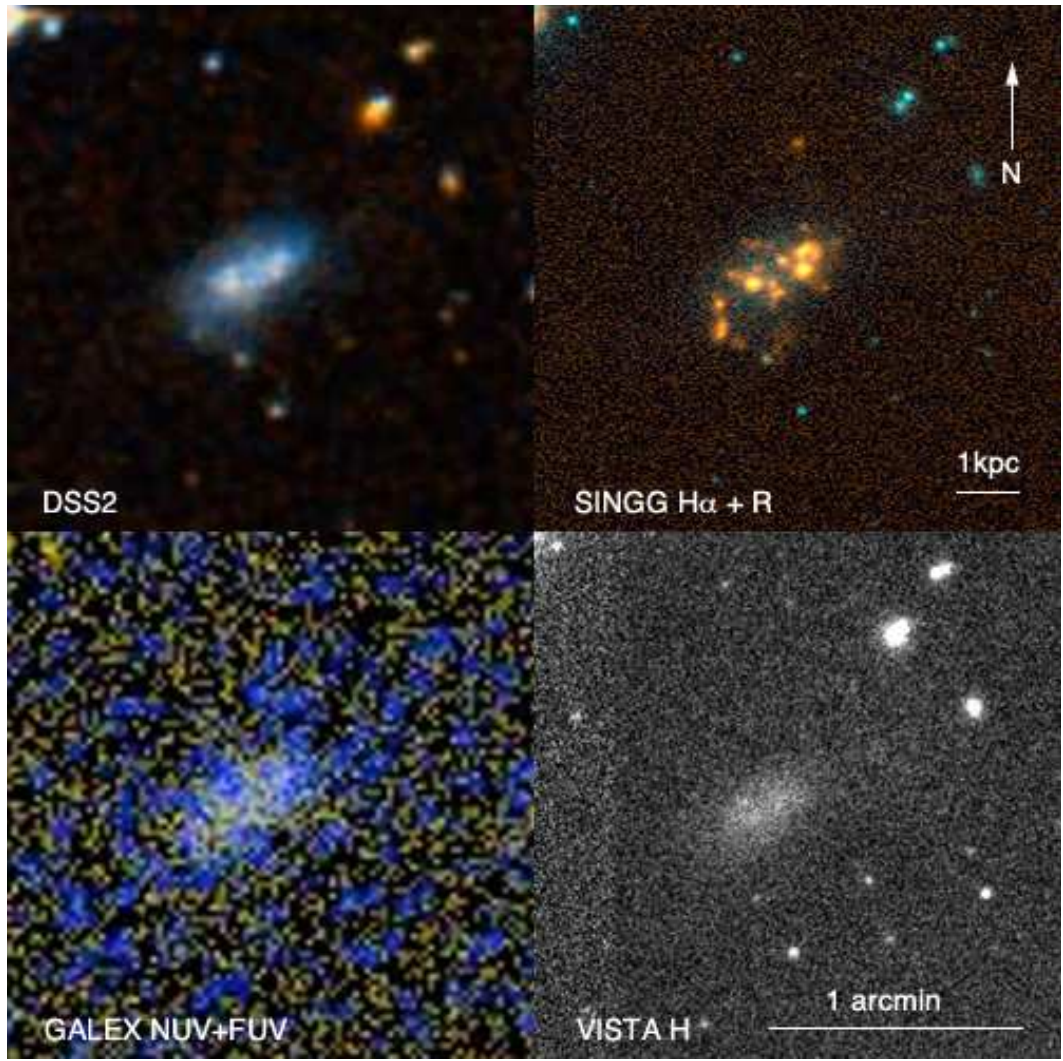}
\caption{J1609-04 in UV, visible, H$\alpha$/R-band and H-band, all to the same scale}\label{fig_3}
\end{figure}

Figure \ref{fig_3} shows the GALEX NUV+FUV, visible, H$_\alpha$ and R-band, and H-band images.  There are several active star forming regions.  The H-band image is from the VISTA Phase III survey (http://www.eso.org/sci/observing/phase3.html) and the H$\alpha$ + R-band is from the SINGG survey \citep{Meurer06}.

Table 2 summarises information available on J1609-04.  Stellar masses were calculated using the methods discussed by \cite{Kirby08}, based on the stellar mass-to-light ratio models of \cite{Bell01}. \cite{Bell01} found that the major contributor to uncertainties in estimating total stellar mass from H- or other IR band magnitudes was the initial mass function.  As this is not well known, we have adopted the same approach as \cite{Kirby08}. Drawing on their more extensive observational base, we assume the same mass-to-light ratio, $\Upsilon_*^H$=1.0$\pm$0.4. 

For the stellar mass and error estimates,  photometry was performed on the VISTA H-band FITS image, using an elliptical aperture large enough to cover the complete galaxy to measure the total H-band flux. The same ellipse was used to measure the noise characteristics in five adjacent, signal free areas (no apparent stars or background galaxies). The errors quoted in Table 2 combine the effect of the (large) mass-to-light ratio uncertainties and the (small) average of the luminosity noise measurements.

\begin{deluxetable}{lll}
\centering
\tabletypesize{\footnotesize}
\tablecaption{Compendium of UV, Optical, Near-IR, and HI Parameters for J1609}\label{table_2}
\tablehead{
\colhead{Parameter} & \colhead{Value} & \colhead{Reference} \\
}
\startdata
R.A. (J2000)				&	16 09 36.8	& NED \\
Dec (J2000)				&	-04 37 12.6	& NED \\
l (deg)						&	7.134450 	&  derived from \cite{Lauberts82}\\
b (deg)						&	32.607155 &  derived from \cite{Lauberts82}\\
Dist (Mpc)						& 	14.82		& NED \\
m$_B$ (mag) 					& 	15.35		& \cite{Doyle05}\\
A$_B$ (mag)					&   	0.862		& \cite{Schlafly11}\\
E$_{B-V}$ (mag.)				& 	0.201		& \cite{Schlafly11} \\
M$_{B,0}$ (mag)				& 	-15.50		&  (derived here) \\
log$_{10}$(L$_B$/L$_\sun$)		& 	8.39			& from NED \\
m$_H$ (mag) 					& 	13.26$\pm$0.06	& from VISTA Phase III imagery$^a$\\
M$_{H,0}$ (mag)				& 	-17.59$\pm$0.06	& from VISTA Phase III imagery\\
FUV flux ($\mu$Jy)				&	361.26$\pm$29.68	& GALEX Data Release 6/7\\
NUV flux ($\mu$Jy)				&	345.87$\pm$17.19	&  GALEX Data Release 6/7\\
Total H~\textsc{i} flux (Jy km  s$^{-1}$)		& 7.2		& \cite{Meyer04}\\
W$_{20}$ (km  s$^{-1}$)			&	111.2		& \cite{Meyer04} \\
W$_{50}$ (km  s$^{-1}$)			&	71.9		         & \cite{Meyer04} \\
log$_{10}(\mathcal{M}_{HI}/\mathcal{M}_{\odot}$)	&	8.55$\pm$0.03		& \cite{Meurer06}$^b$\\
log$_{10}(\mathcal{M}_{\ast}/{\mathcal{M}}_{\odot}$)	&	8.37$\pm$0.22 	& from VISTA Phase III imagery$^c$\\
M$_{\rm H\textsc{i}}$/L$_B$		&	1.45$\pm$0.10			& (derived here)\\
$\Delta$ tidal index				&	-2.9			& \cite{Nicholls11}\\
log$_{10}$(ii) isolation index		&	2.38			& \cite{Karachentsev11}\\
\enddata
\tablecomments{(a) The H-band magnitude was calculated from the VISTA image using the standard equation \\
m$_H$ = -2.5 log$_{10}$(flux) + photozp\\
(b) Re-calculated using the flow-corrected distance taking into account the Virgo Cluster, the Great Attractor and the Shapley concentration.  \\
(c) The major part or the error ($\pm$ 0.2) arises from uncertainties in determining the mass-to-light ratio from the H-band luminosity. See above.}\\
\end{deluxetable}

\FloatBarrier 

\section{Observations and Data reduction}

\begin{figure*}[htpb]
\centering
\includegraphics[width=0.454\hsize]{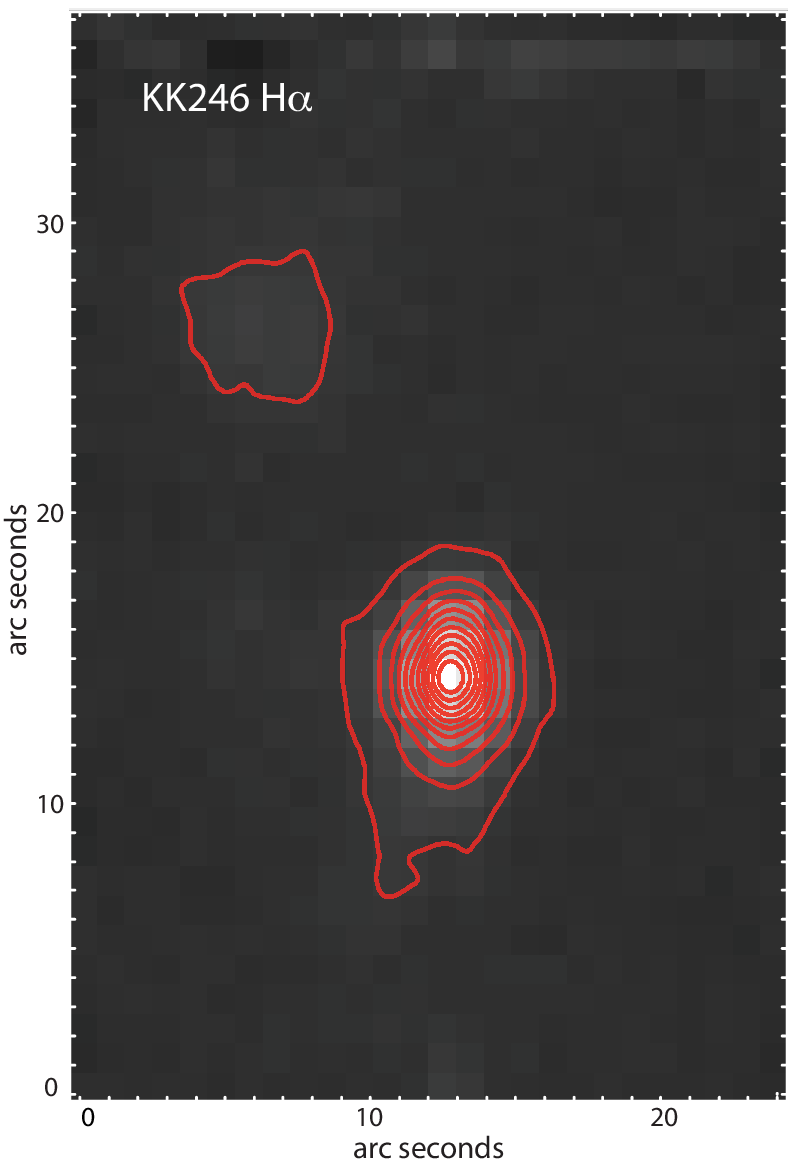}
\includegraphics[width=0.46\hsize]{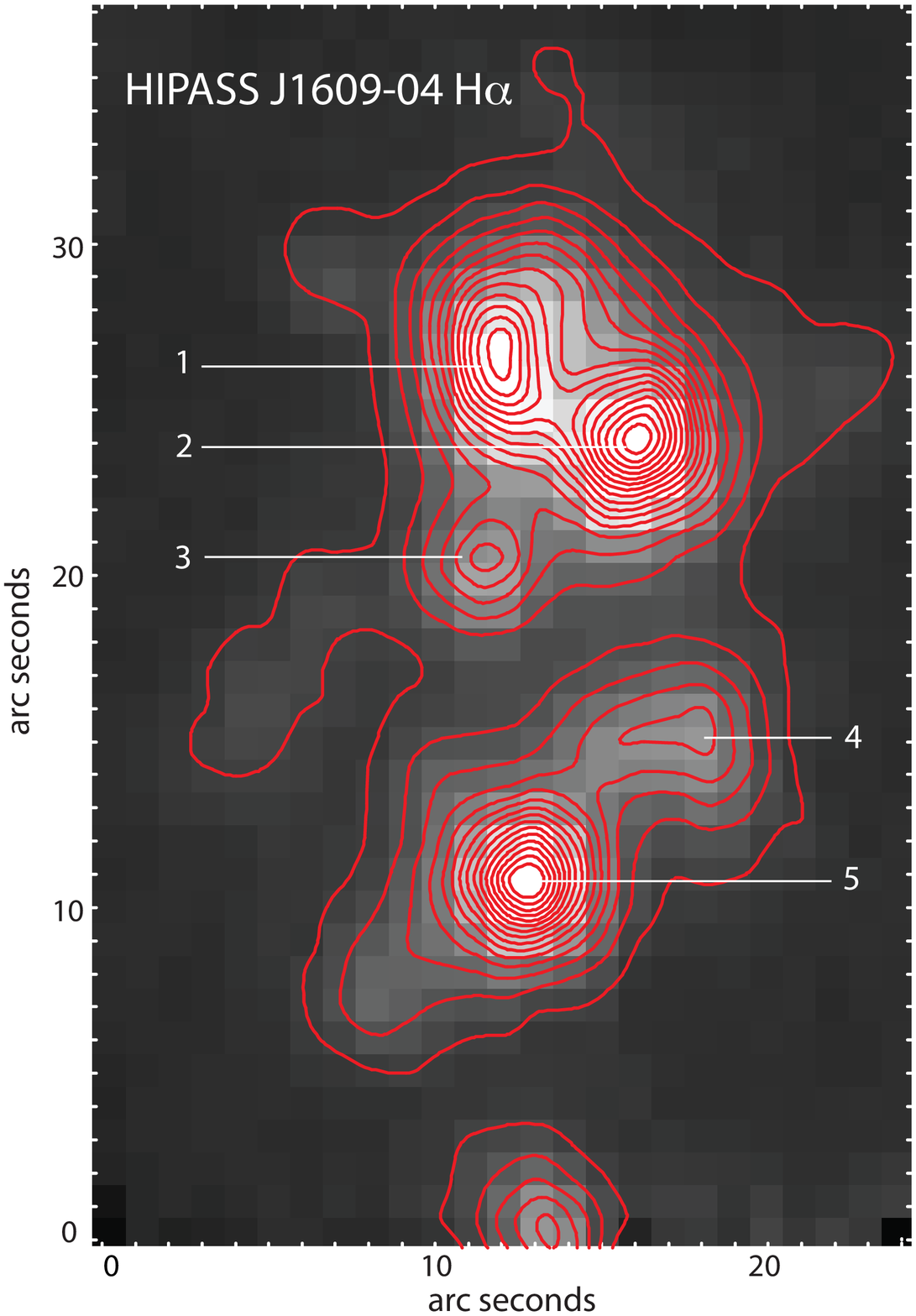}
\caption{Left Panel: KK246 H$\alpha$ WiFeS IFU image with contours (logarithmic stretch, 6.2e-18 to 1.76e-16 erg cm$^{-2}$ sec$^{-1}$ \AA$^{-1}$, at 0.125 dex intervals). 
Right Panel: J1609-04 H$\alpha$ WiFeS IFU image with contours (logarithmic stretch, 1.0e-18 to 7.2e-16 erg cm$^{-2}$ sec$^{-1}$ \AA$^{-1}$, at 0.25 dex intervals). Numbers indicate the locations of the individual \HII regions used in the analysis (Tables \ref{table_3} to \ref{table_7}).}\label{fig_4}
\end{figure*}

We observed both target galaxies using the WiFeS IFU spectrograph \citep{Dopita07, Dopita10} on the ANU 2.3 meter telescope at Siding Spring Observatory. The WiFeS spectrograph is ideal for objects of this size, with a field of view of 25$\times$38 arc seconds.
Data were obtained on 23--26 August 2011, under clear skies with seeing 1.6 - 2.4 arcsec, at resolutions of R=3000 ($\lambda < 5600$\AA) and 7000 (5600 $< \lambda <$ 7000\AA), with signal-to-noise ratios typically 50 to 100:1 for the brighter lines.  The total on-source integration time was 3.5 hours for KK246 and 1 hour for J1609-04, respectively. Figure \ref{fig_4} shows the 3\AA-wide H$\alpha$ slices (6571 to 6574\AA) from the data cubes, with overlaid flux contours obtained using IRAF/STSDAS/newcont.  

Observed, reddening-corrected  line fluxes are given in Table \ref{table_3}.  Uncorrected data are shown in Table \ref{table_4}. The data were reduced using the revised WiFeS Python pipeline \citep{Childress13}. This involves steps generally similar to those described in \cite{Dopita10} for the older pipeline: bias modelling and subtraction, arc line identification and wavelength calibration, cosmic ray removal, sky-line subtraction using nod-and-shuffle, initial data cube construction and atmospheric dispersion correction, standard flux star calibration, telluric correction, assembly into the final data cubes and combination of multiple cubes into a final object data cube. Spectral sampling was undertaken using a 3 arc second radius circular spatial area centered on each \HII region, through the full wavelength range of the data cube, to obtain spectra for each region.Lines fluxes were measured from these spectra using IRAF/splot, and flux de-reddening was performed on the raw flux data using the dust models from \cite{Fischera05}, using a relative extinction curve with R$^A_V$=4.3, where R$^A_V$ = A$_V$/(E$_{B-V}$) and A$_V$ is the V-band extinction.  See also the discussion in \citet[Appendix 1]{Vogt13}. We used an initial Balmer decrement ratio of 2.82 for H$\alpha$/H$\beta$, then adjusted the apparent Balmer ratios by varying the value of A$_V$ for the best fit to the H$\gamma$/H$\beta$ ratio, using the ratio H$\delta$/H$\beta$ as a check, fitting to the \cite{Storey98} Case B Balmer ratios.

The raw fluxes are listed in Table \ref{table_4}, and the de-reddening  fits for the Balmer ratios of H$\alpha$ though H$\delta$ in Table \ref{table_5}.  The fit for region 3 of J1609-04 is not particularly satisfactory, possibly due to differential contamination of the sampled area from regions 1 and 2 in the red and blue channels.  Other fits are good, and excellent in the case of KK246.  The low value of the extinction, A$_V$=0.032, for KK246, implies very low dust content. For clarity the error estimates are omitted from Table \ref{table_4}, but they may be calculated from those quoted in Table \ref{table_3}.

\FloatBarrier 
\begin{landscape}
\begin{table}[htbp]
  \centering
  \caption{Measured line fluxes for the \HII  regions in KK246 and HIPASS J1609-04, corrected for reddening, in units of erg sec$^{-1}$cm$^{-2}$\AA$^{-1}$. Fluxes were measured using a 3 arc second radius sample centered on each \HII region.}
\vskip 1em
  {\small
    \begin{tabular}{l|l|l|l|l|l|l}
   \hline \hline
    Line & KK246 &   J1609(1) &   J1609(2) &   J1609(3) &   J1609(4) &   J1609(5)  \\
\hline
SII 6731 & (2.26$\pm$0.16)E-16 & (8.54$\pm$0.23)E-15 & (1.18$\pm$0.02)E-15 & (5.28$\pm$0.21)E-16 & (4.55$\pm$0.18)E-16 & (7.86$\pm$0.02)E-15 \\
 &  &  &  &  &  & \\
SII 6716 & (3.34$\pm$0.15)E-16 & (1.21$\pm$0.24)E-15 & (1.67$\pm$0.02)E-15 & (7.78$\pm$0.21)E-16 & (6.22$\pm$0.18)E-16 & (1.15$\pm$0.02)E-15 \\
 &  &  &  &  &  & \\
NII 6584 & (1.65$\pm$0.17)E-16 & (4.65$\pm$0.17)E-16 & (6.67$\pm$0.19)E-16 & (3.41$\pm$0.19)E-16 & (2.42$\pm$0.18)E-16 & (5.92$\pm$0.19)E-16 \\
 &  &  &  &  &  & \\
H$\alpha$ 6563 & (6.35$\pm$0.02)E-15 & (1.33$\pm$0.01)E-14 & (1.60$\pm$0.01)E-14 & (6.31$\pm$0.02)E-15 & (5.92$\pm$0.03)E-15 & (1.18$\pm$0.00)E-14 \\
 &  &  &  &  &  & \\
OIII 5007 & (5.65$\pm$0.03)E-15 & (1.38$\pm$0.01)E-14 & (1.50$\pm$0.01)E-14 & (5.65$\pm$0.02)E-15 & (5.82$\pm$0.02)E-15 & (1.34$\pm$0.00)E-14 \\
 &  &  &  &  &  & \\
OIII 4959 & (2.02$\pm$0.03)E-15 & (4.63$\pm$0.03)E-15 & (5.12$\pm$0.02)E-15 & (1.88$\pm$0.02)E-15 & (1.93$\pm$0.02)E-15 & (4.62$\pm$0.02)E-15 \\
 &  &  &  &  &  & \\
H$\beta$ 4861 & (2.24$\pm$0.04)E-15 & (4.76$\pm$0.11)E-15 & (5.59$\pm$0.03)E-15 & (2.27$\pm$0.02)E-15 & (2.10$\pm$0.02)E-15 & (4.20$\pm$0.02)E-15 \\
 &  &  &  &  &  & \\
OIII 4363 & - & - & (1.34$\pm$0.15)E-16 & - & - & (2.90$\pm$0.29)E-16 \\
 &  &  &  &  &  & \\
H$\gamma$ 4340 & (1.06$\pm$0.04)E-15 & (2.26$\pm$0.03)E-15 & (2.70$\pm$0.03)E-15 & (9.65$\pm$0.26)E-16 & (9.93$\pm$0.28)E-16 & (1.97$\pm$0.03)E-15 \\
 &  &  &  &  &  & \\
 H$\delta$ 4102 & (5.82$\pm$0.06)E-16 & (1.23$\pm$0.03)E-15 & (1.48$\pm$0.04)E-15 & (4.92$\pm$0.36)E-16 & (5.19$\pm$0.38)E-16 & (9.83$\pm$0.34)E-16 \\
 &  &  &  &  &  & \\
OII 3729 & (2.50$\pm$0.15)E-15 & (7.54$\pm$0.07)E-15 & (9.49$\pm$0.09)E-15 & (3.67$\pm$0.10)E-15 & (5.30$\pm$0.12)E-15 & (6.44$\pm$0.07)E-15 \\
 &  &  &  &  &  & \\
OII 3726 & (1.37$\pm$0.15)E-15 & (5.95$\pm$0.20)E-15 & (8.50$\pm$0.09)E-15 & (3.54$\pm$0.10)E-15 & (3.92$\pm$0.12)E-15 & (5.93$\pm$0.07)E-15 \\
          &     &       &       &       &       &      \\
    \hline \hline
    \end{tabular}
    }
  \label{table_3}
\end{table}
\end{landscape}

\begin{table}[htbp]
  \centering
  \caption{Uncorrected fluxes  (erg sec$^{-1}$cm$^{-2}$\AA$^{-1}$). For errors see previous table.}
\vskip 1em
  {\normalsize
\begin{tabular}{lrrrrrr}
   \hline \hline
Line   	  & KK246	 & J1609(1) & J1609(2) & J1609(3) & J1609(4) & J1609(5)  \\ \hline
		  &      	 &       	&		   &       	  &          &     \\
SII 6731  & 2.29E-16 & 1.39E-15 & 1.63E-15 & 7.55E-16 & 7.10E-16 & 1.24E-15 \\
SII 6716  & 3.38E-16 & 1.96E-15 & 2.31E-15 & 1.11E-15 & 9.66E-16 & 1.80E-15 \\
NII 6584  & 1.67E-16 & 7.35E-16 & 9.08E-16 & 4.78E-16 & 3.68E-16 & 9.09E-16 \\
H$\alpha$    	  & 6.42E-15 & 2.10E-14 & 2.17E-14 & 8.82E-15 & 8.96E-15 & 1.81E-14 \\
OIII 5007 & 5.66E-15 & 1.45E-14 & 1.55E-14 & 5.86E-15 & 6.09E-15 & 1.41E-14 \\
OIII 4959 & 2.02E-15 & 4.79E-15 & 5.24E-15 & 1.93E-15 & 1.99E-15 & 4.76E-15 \\
H$\beta$    	  & 2.24E-15 & 4.76E-15 & 5.59E-15 & 2.27E-15 & 2.10E-15 & 4.20E-15 \\
OIII 4363 &     -     &    -     & 1.18E-16 &    -    &     -    & 2.42E-16 \\
H$\gamma$  	  & 1.05E-15 & 1.84E-15 & 2.35E-15 & 8.30E-16 & 8.24E-16 & 1.63E-15 \\
H$\delta$ 	  & 5.77E-16 & 9.01E-16 & 1.20E-15 & 3.91E-16 & 3.91E-16 & 7.35E-16 \\
OII 3729  & 2.29E-15 & 4.63E-15 & 6.84E-15 & 2.56E-15 & 3.40E-15 & 4.08E-15 \\
OII 3726  & 1.37E-15 & 3.65E-15 & 6.12E-15 & 2.47E-15 & 2.51E-15 & 3.75E-15 \\
		  &          &          &          &          &          &          \\
\hline \hline
\end{tabular}
}
  \label{table_4}
\end{table}

\begin{table}[htbp]
  \centering
  \caption{Corrected Balmer line ratios and corresponding A$_v$ values}
\vskip 1em
  {\normalsize
\begin{tabular}{rrrrrrr}
   \hline \hline
Object & KK246 & J1609(1) & J1609(2) & J1609(3) & J1609(4) & J1609(5) \\ \hline 
Av    & 0.032 & 1.250 & 0.840 & 0.920 & 1.140 & 1.170 \\ \hline
H$\alpha$/H$\beta$ & 2.833 & 2.802 & 2.861 & 2.782 & 2.820 & 2.818 \\
H$\gamma$/H$\beta$ & 0.471 & 0.475 & 0.483 & 0.425 & 0.473 & 0.470 \\
H$\delta$/H$\beta$ & 0.260 & 0.258 & 0.265 & 0.217 & 0.247 & 0.234 \\
\hline \hline
\end{tabular}
}
  \label{table_5}
\end{table}

\FloatBarrier 

\section{Spectra}
\begin{figure}[htpb]
\centering
\includegraphics[width=0.85\hsize]{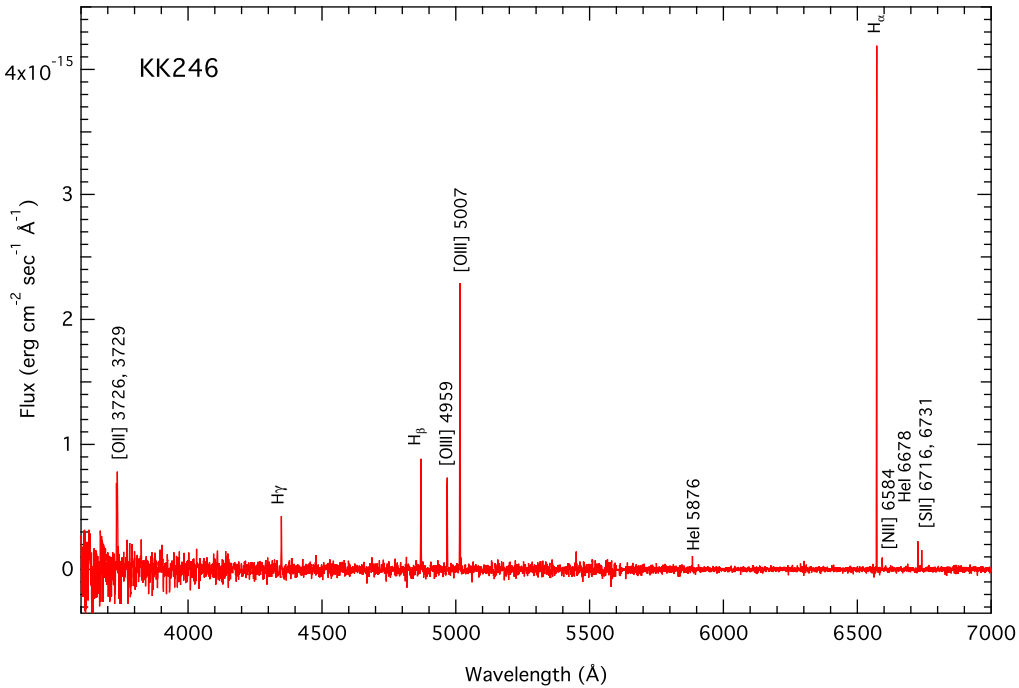}
\caption{Flux-calibrated spectrum of KK246 with principal atomic lines labelled.}\label{fig_5}
\end{figure}

Figure \ref{fig_5} shows the flux-calibrated spectrum of KK246 in the range $3600\AA<\lambda<7000\AA$ derived from the data cubes using spatial pixels within a 6 arc second diameter aperture.  The principle atomic lines are: [OII] 3726,27\AA, H$_\gamma$, H$_\beta$, [OIII] 4959\AA, [OIII] 5007\AA, HeI 5876, H$_\alpha$, [NII] 6584\AA, HeI 6678, and [SII] 6716,31\AA.  It is noteworthy that the intensity of the [OIII] 4959\AA \ line is less than the H$_\beta$, the intensity of the [OIII] 5007\AA \ line is considerably less than the H$_\alpha$, and there is no evidence of the [OIII] 4363\AA \ auroral line.  This suggests a reasonably low metallicity, and possibly the tail-end of an episode of star formation, and hence relatively low ionization. No stellar continuum is apparent.

\FloatBarrier

\begin{figure}[htpb]
\centering
\includegraphics[width=0.85\hsize]{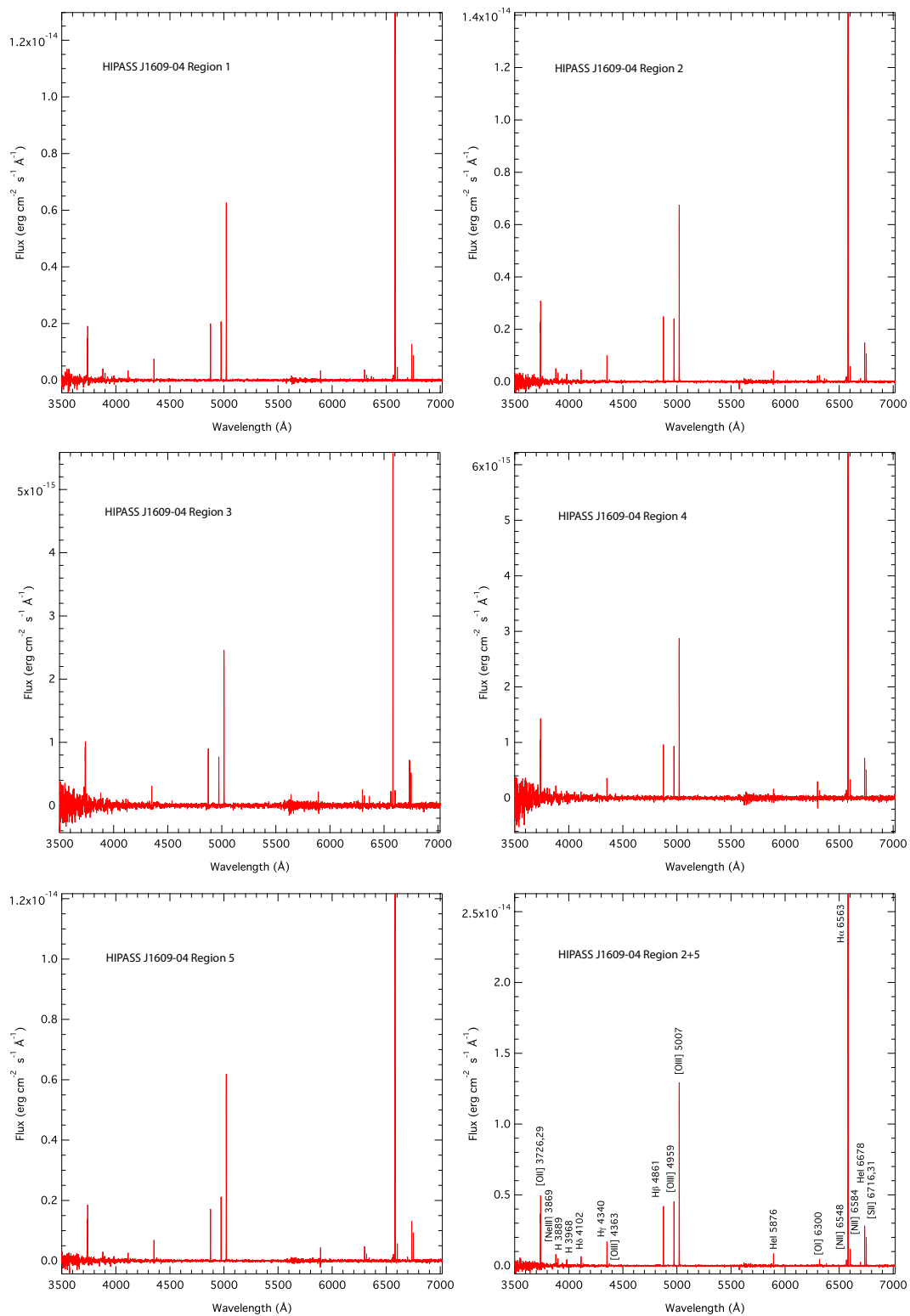}
\caption{Spectra of five individual HII regions in HIPASS J1609-04, and a combined spectrum from regions (2) and (5), with line identification.}\label{fig_6}
\end{figure}

Five separate, reasonably bright, star forming regions were observed in HIPASS \mbox{J1609-04}, and spectra for each of these were obtained. 
The regions are identified in Figure \ref{fig_4}, right panel. The spectra are shown in Figure \ref{fig_6}.  The auroral line is apparent in the two brightest 
regions, 2 and 5.  The bottom right panel of Figure \ref{fig_6} is a composite spectrum of regions 2 and 5, with the main spectral lines labeled.

\subsection{Comments on spectra}
The spectra of both KK246 and J1609-04 show, at most, only faint stellar continua, and as a result measurement of equivalent widths of the spectral lines is unreliable.  The brighter \HII regions (2) and (5) in J1609-04 exhibit faint broad stellar absorption underlying the H$\beta$, $\gamma$ and $\delta$ nebular emission lines, but the stellar continuum is faint. Measurement noise makes it difficult to estimate the equivalent widths of the absorption features, but they have been taken into account in estimating emission line fluxes. In addition, our reduction pipeline does not currently allow accurate rendering of the absolute continuum for faint continua at the short wavelength end of the blue spectrum, below 4200\AA. 

Equivalent widths for emission lines are similarly problematic. The low stellar continuum creates large EW values for the bright lines , e.g. EW [OIII5007] = 135 for J1609-04(5), with reliable errors difficult to estimate.  For the UV [OII] lines, the measurement noise is sufficiently large that EW estimates are meaningless. For this reason we have not included EW measurements in this paper.
\FloatBarrier

\section{Nebular metallicities}

The most important insight into the evolutionary stage of the two objects is provided by their (nebular) metallicities. This tells us about the enrichment of the star-forming regions by previous generations of stars, and the degree to which the heavier elements are retained by the gravitational potential of the galaxies.  In general, the mass-metallicity relation shows that there is a trend to lower metallicities for lower mass galaxies \citep[e.g.,][]{Tremonti04}.  The primary finding of this paper relates to metallicity measurements.

\begin{figure}[htpb]
\centering
\includegraphics[width=0.85\hsize]{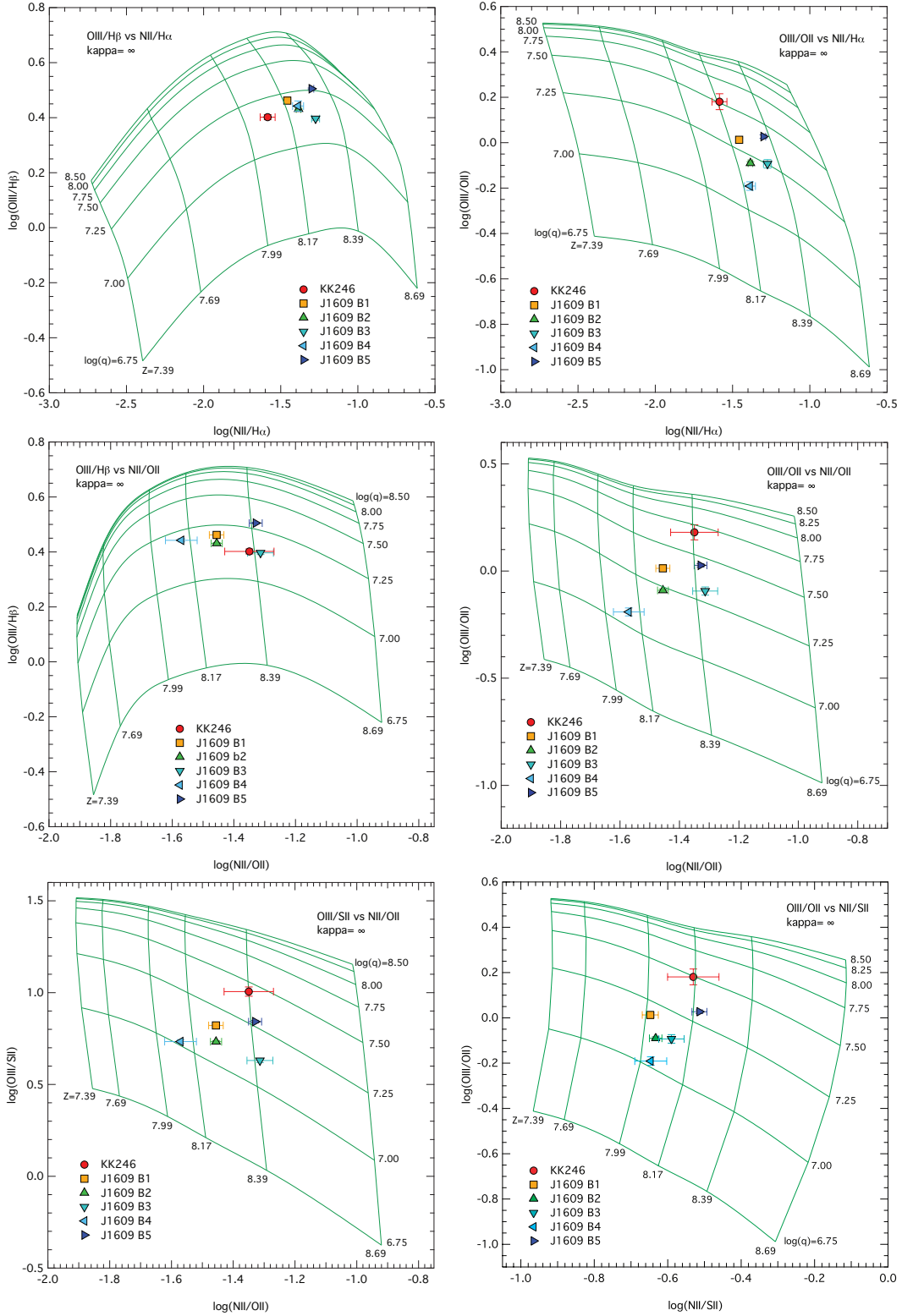}
\caption{Strong line diagnostic grids, showing flux ratios for KK246 and the 5 regions in J1609-04 (grids computed using data from \cite{Dopita13}).}\label{fig_7}
\end{figure}

\begin{figure}[htpb]
\centering
\includegraphics[width=0.85\hsize]{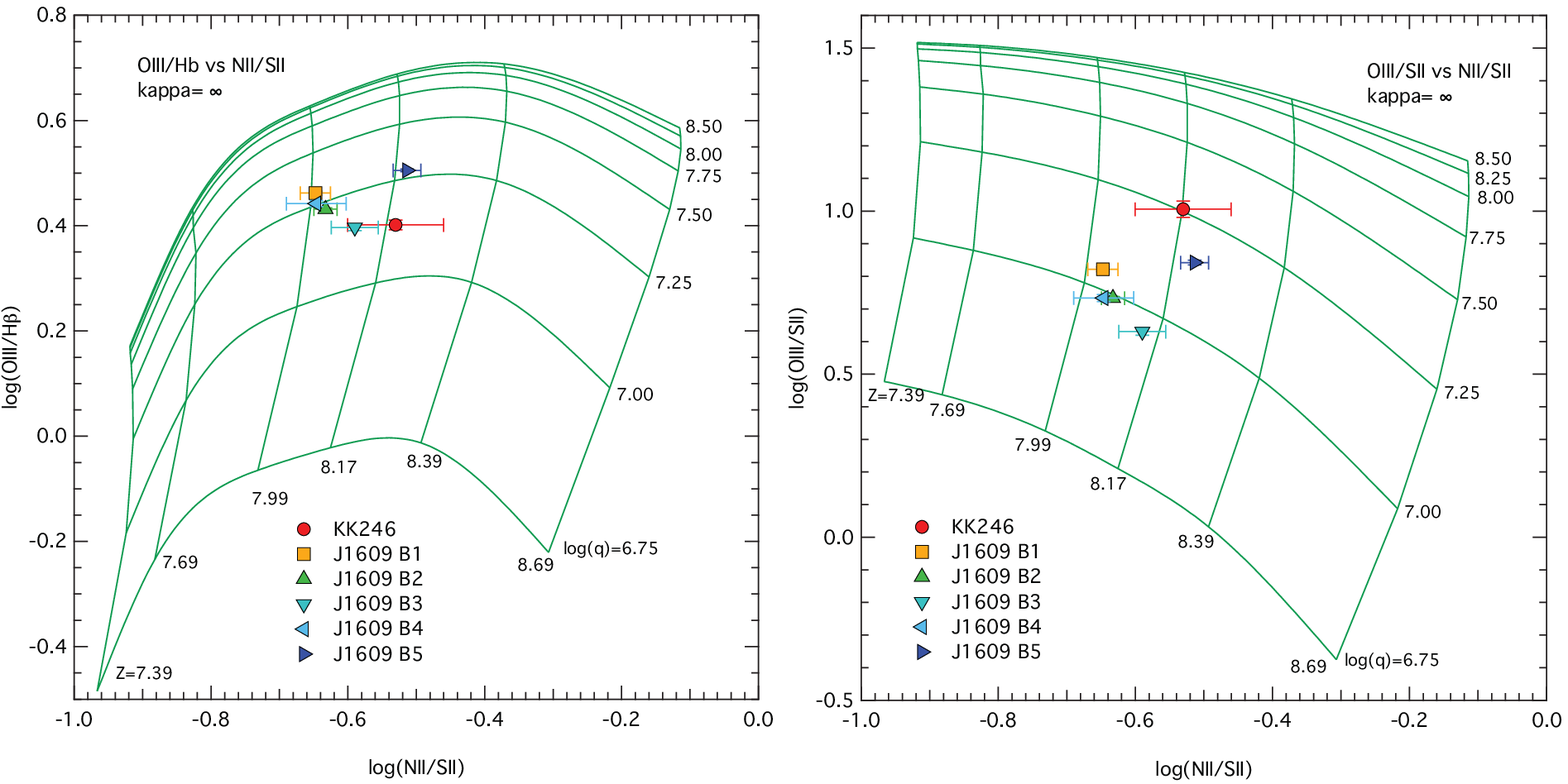}
\caption{Strong line diagnostic grids, continued, showing flux ratios for KK246 and the 5 regions in J1609-04 (grids computed using data from \cite{Dopita13})}\label{fig_8}
\end{figure}

A key method for determining nebular temperatures and metallicities is the electron temperature, or $T_e$ method.  This is possible only if one of the ``auroral'' lines is observed, usually [O III] 4363\AA.  This line was detected in only two of the five regions in J1609-04, but was absent in the spectrum of KK246, so we have used strong line methods to determine metallicities \citep[e.g.,][]{Dopita13, Kewley02, Kewley08}. \citet{Lopez12} evaluated the reliability of several of the conventional strong line methods in reproducing theoretically constructed emission line data.  They found that some of the commonly used techniques cannot be relied on to give consistently accurate results for nebular metallicity, and that the $T_e$ method gave consistently lower metallicities than the (reliable) strong line methods, by $\sim$0.2-0.25 dex.  Subsequently, \cite{Nicholls12,Nicholls13} showed that non-equilibrium ($\kappa$) electron distributions can affect the diagnostic results.  This, and the advent of much improved atomic data, led us \citep{Nicholls13,Dopita13} to revise the strong line diagnostic methods and the photoionization modelling code.  The strong line diagnostics used here are derived from the newly revised version of the Mappings photoionization modelling code (version IV)   \citep{Dopita13}, resulting in substantially more consistent metallicity values, as demonstrated in Table \ref{table_6}, which shows the results of that analysis.  The diagnostic grids on which the analysis is based are shown in Figures \ref{fig_7} and \ref{fig_8}.

Earlier strong line methods did not solve explicitly for the ionization parameter, log(q), and thus did not take fully into account the ionization gradients present in \HII regions \citep[see, e.g.,][]{Dopita13}. In the analysis used in Mappings IV, we solve separately for the metallicity Z (= 12+log(O/H)) and ionization parameter log(q), thus avoiding most of the degeneracies present in older strong line diagnostic methods. The diagnostics chosen here are also relatively insensitive to non-equilibrium ($\kappa$) electron energy distributions, especially for values of $\kappa > $ 20, i.e., values which give the best fit in resolving the abundance discrepancy \citep{Stasinska04, Nicholls12}.

\FloatBarrier

\begin{deluxetable}{lllllll}
\centering
\tabletypesize{\scriptsize}
\tablecaption{Metallicity and ionization parameter values (see section 6.2)\label{table_6}}
\tablehead{
\colhead{ } & \colhead{KK246} & \colhead{J1609(1)} & \colhead{J1609(2)} & \colhead{J1609(3)} & \colhead{J1609(4)} & \colhead{J1609(5)} 
}
\startdata
$\kappa = \infty$ &       &       &       &       &       &  \\
   \hline  \\
NII/SII vs OIII/SII &       &       &       &       &       &  \\
 z  & 8.172 & 8.027 & 8.066 & 8.144 & 8.038 & 8.211 \\
 log(q) & 7.256 & 7.051 & 6.993 & 6.973 & 6.990 & 7.157 \\
NII/SII vs OIII/Hb &       &       &       &       &       &  \\
 z  & 8.187 & 8.00 & 8.031 & 8.111 & 8.001 & 8.195 \\
 log(q) & 7.142 & 7.304 & 7.242 & 7.154 & 7.257  & 7.294 \\
NII/OII vs OIII/OII &       &       &       &       &       &  \\
 z  & 8.373 & 8.250 & 8.284 & 8.398 & 8.079 & 8.408 \\
 log(q) & 7.637 & 7.316 & 7.242 & 7.277 & 7.025 & 7.449 \\
NII/OII vs OIII/SII &       &       &       &       &       &  \\
 z  & 8.367 & 8.240 & 8.273 & 8.388 & 8.076 & 8.401 \\
 log(q) & 7.420 & 7.156 & 7.101 & 7.109 & 7.002 & 7.281 \\
NII/SII vs OIII/OII &       &       &       &       &       &  \\
 z  & 8.162 & 8.008 & 8.042 & 8.112 & 8.035 & 8.194 \\
 log(q) & 7.474 & 7.184 & 7.123 & 7.132 & 7.005 & 7.318 \\
NII/OII vs OIII/Hb &       &       &       &       &       &  \\
 z  & 8.351 & 8.244 & 8.280 & 8.390 & 8.102 & 8.403 \\
 log(q) & 7.134 & 7.209 & 7.166 & 7.130 & 7.212 & 7.296 \\
NII/Ha vs OIII/Hb &       &       &       &       &       &  \\
 z  & 8.095 & 8.196 & 8.244 & 8.303 & 8.245 & 8.334 \\
 log(q) & 6.560 & 6.754 & 7.181 & 7.012 & 7.194 & 7.212\\
NII/Ha vs OIII/OII &       &       &       &       &       &  \\
 z  & 8.176 & 8.218 & 8.252 & 8.333 & 8.203 & 8.361 \\
 log(q) & 7.482 & 7.294 & 7.221 & 7.234 & 7.088 & 7.420 \\
 
%\hline%     &       &       &       &       &       &  \\
\rule{0pt}{4ex}mean z & 8.24 $\pm$ 0.11 & 8.15 $\pm$ 0.11 & 8.18 $\pm$ 0.12 & 8.27 $\pm$ 0.13 & 8.10 $\pm$ 0.09 & 8.31 $\pm$ 0.10 \\
mean log(q) & 7.26 $\pm$ 0.33 & 7.16 $\pm$ 0.19 & 7.16 $\pm$ 0.09 & 7.13 $\pm$ 0.10 & 7.10 $\pm$ 0.11 & 7.30 $\pm$ 0.10 \\
 %\hline%    &       &       &       &       &       &  \\
\rule{0pt}{4ex}adopted mean z & 8.17 $\pm$ 0.01 & 8.01 $\pm$ 0.01 & 8.05 $\pm$ 0.02 & 8.12 $\pm$ 0.02 & 8.03 $\pm$ 0.02 & 8.20 $\pm$ 0.01  \\ 
adopted  mean log(q) & 7.29 $\pm$ 0.17 & 7.18 $\pm$ 0.13 & 7.11 $\pm$ 0.12 & 7.09 $\pm$ 0.10 & 7.08 $\pm$ 0.15 & 7.26 $\pm$ 0.09 \\ 
          &       &       &       &       &       &  \\
          \hline% \\
    $\kappa$ = 50 &       &       &       &       &       &  \\
   \hline  \\
   
mean z & 8.25 $\pm$ 0.10 & 8.17 $\pm$ 0.13 & 8.20 $\pm$ 0.13 & 8.29 $\pm$ 0.14 & 8.11 $\pm$ 0.12 & 8.32 $\pm$ 0.11 \\ 
mean log(q) & 7.36 $\pm$ 0.15 & 7.24 $\pm$ 0.13 & 7.20 $\pm$ 0.11 & 7.18 $\pm$ 0.10 & 7.15 $\pm$ 0.16 & 7.33 $\pm$ 0.10 \\
%       &       &       &       &       &       &  \\
\rule{0pt}{4ex}adopted mean z & 8.17 $\pm$ 0.01 & 8.02 $\pm$ 0.02 & 8.05 $\pm$ 0.02 & 8.12 $\pm$ 0.02 & 8.03 $\pm$ 0.02 & 8.20 $\pm$ 0.01  \\ 
adopted  mean log(q) & 7.32 $\pm$ 0.12 & 7.25 $\pm$ 0.20 & 7.18 $\pm$ 0.18 & 7.13 $\pm$ 0.13 & 7.16 $\pm$ 0.23 & 7.30 $\pm$ 0.13 \\ 
          &       &       &       &       &       &  \\
          \hline% \\
    $\kappa$ = 20 &       &       &       &       &       &  \\
 \hline  \\
mean z & 8.28 $\pm$ 0.09 & 8.18 $\pm$ 0.15 & 8.19 $\pm$ 0.14 & 8.26 $\pm$ 0.15 & 8.08 $\pm$ 0.13 & 8.29 $\pm$ 0.11 \\
mean log(q) & 7.43 $\pm$ 0.09 & 7.37 $\pm$ 0.23 & 7.28 $\pm$ 0.20 & 7.24 $\pm$ 0.16 & 7.19 $\pm$ 0.28 & 7.38 $\pm$ 0.19 \\
 %      &       &       &       &       &       &  \\
\rule{0pt}{4ex}adopted mean z & 8.17 $\pm$ 0.01 & 8.02 $\pm$ 0.02 & 8.05 $\pm$ 0.03 & 8.12 $\pm$ 0.04 & 8.03 $\pm$ 0.03 & 8.21 $\pm$ 0.02  \\ 
adopted  mean log(q) & 7.39 $\pm$ 0.08 & 7.40 $\pm$ 0.40 & 7.29 $\pm$ 0.31 & 7.23 $\pm$ 0.22 & 7.29 $\pm$ 0.39 & 7.40 $\pm$ 0.25 \\  
\enddata
\end{deluxetable}

\subsection{Direct $T_e$ method metallicity}

For the two brightest regions in J1609-04 (2 and 5), the [OIII] auroral line $\lambda$4363 is evident. For these we can measure the metallicity using the direct $T_e$ method.  To improve the signal-to-noise ratio, we have combined the spectra of the two regions. Using the models presented in \cite{Dopita13, Nicholls13}, we calculate the [O III] electron temperature from the combined spectra, as: $T_e$ = 12,806$\pm$556 K. Using the methods described in \cite{Izotov06}, the oxygen ([OIII]+O[II]) metallicity calculated from the electron temperature of the combination of regions 2 and 5 of J1609-04 is 7.96$\pm$0.06. As is characteristic of metallicities derived conventionally from the $T_e$ method \citep{Lopez12}, this value is lower by $\sim$ 0.2 dex than the value determined using the latest strong line methods \citep{Dopita13}.

\subsection{Metallicity results (strong line)}

Table \ref{table_6} shows the complete line ratios for $\kappa = \infty$, with two sets of averages. The first set takes the means over all strong line diagnostics. The second, significantly more precise, uses only the mean values of diagnostic ratios using [NII]/[SII]. The latter diagnostic ratios accurately return input values from test spectra generated by the Mappings IV code, whereas the diagnostic ratios involving [NII]/[OII] return significantly higher values, in some cases.  It is not clear yet whether this indicates a problem with the diagnostic itself, or, more likely, whether the interpolation process \citep{Dopita13} is at fault. Further work is necessary.  In any case, as shown in Figures 7 and 8, the errors in observed flux ratios are significantly greater for ratios involving [OII].  For this reason we adopt values for Z and log(q) using only diagnostics involving [NII]/[SII].  Other ratios including [NII]/[Ha] tend to confirm the [NII]/[SII] values.

As shown in Table \ref{table_6}, the metallicity for KK246, assuming an equilibrium electron energy distribution, is 8.17$\pm$0.01, and for  the \HII regions in J1609-04, the metallicity values span a range between 8.02 and 8.20. For a non-equilibrium $\kappa$ electron energy distributions with $\kappa$ = 50 and 20, the metallicities are virtually identical, a consequence of choosing strong line diagnostics that are not especially sensitive to $\kappa$, and which therefore result in very consistent metallicities.

The interpretation of the new diagnostic ratios used here is explored in depth by \cite{Dopita13}, but there is scope for further work.  In the meantime, especially for low metallicity \HII regions, diagnostics involving [NII]/[SII] appear to offer very consistent results.

\section {Comparison with similar samples} 

\subsection{log(N/O) values}

One of the more important parameters in understanding galactic evolution is the nitrogen metallicity, and in particular, the ratio of nitrogen to oxygen.  The observations reported here include good measurements of both [NII] and [OII], allowing us to explore the values of log(N/O) for each \HII region.  To calculate the value of N/O from [NII] and [OII] line fluxes, we use empirical formulae from \cite{Izotov06}, equations (3) and (6).  These reduce to:

\begin{equation}\label{eq1}
log\left(\frac{N}{O}\right) = log\left(\frac{NII~6584+6548}{OII~3726+3729}\right) + 0.273 - 0.726/T_{e4} +0.007*T_{e4} - 0.02*log(T_{e4}) ,
\end{equation}
where T$_{e4}$ is the electron temperature in units of 10,000K. This equation differs only by a small constant offset (0.033) from that quoted by \citet[equation 9]{Pagel92}. We assume the same electron temperature for  OII and NII (reasonable, as they both arise primarily from the outer parts of the \HII region), and further, that N$^+$/O$^+$=N/O, following \cite{Pilyugin10} and others. The errors from these assumptions are likely to be of the same order as the measurement uncertainties. The results are shown in Table \ref{table_7}.

\begin{table}[htbp]
  \centering
  \caption{log(N/O) calculated from [NII]/[OII] flux ratios using equation \eqref{eq1}, for T$_e$ in units of 10,000K. (The measurement errors apply to all ratios.)}
    \vskip 1em
    \begin{tabular}{l|rrrrrr}
    \hline \hline
          & KK246 & J1609(1) & J1609(2) & J1609(3) & J1609(4) & J1609(5) \\ \hline
    log[NII]/[OII] & -1.352 & -1.303 & -1.249 & -1.325 & -1.420 & -1.320 \\
    $\pm$ error & 0.156 & 0.036 & 0.028 & 0.042 & 0.082 & 0.021 \\
    log(N/O) T$_{e4}$=1 & -1.80 & -1.75 & -1.70 & -1.77 & -1.86 & -1.77 \\
    log(N/O) T$_{e4}$=1.25 & -1.65 & -1.60 & -1.55 & -1.63 & -1.72 & -1.62 \\ \hline
    \end{tabular}
  \label{table_7}
\end{table}

The values of log(N/O) for these regions are uniformly low, and similar to  the value adopted by \cite{Lebouteiller13} for 1Zw18, -1.61$\pm$0.10.

\begin{figure}[htpb]
\centering
\includegraphics[width=0.75\hsize]{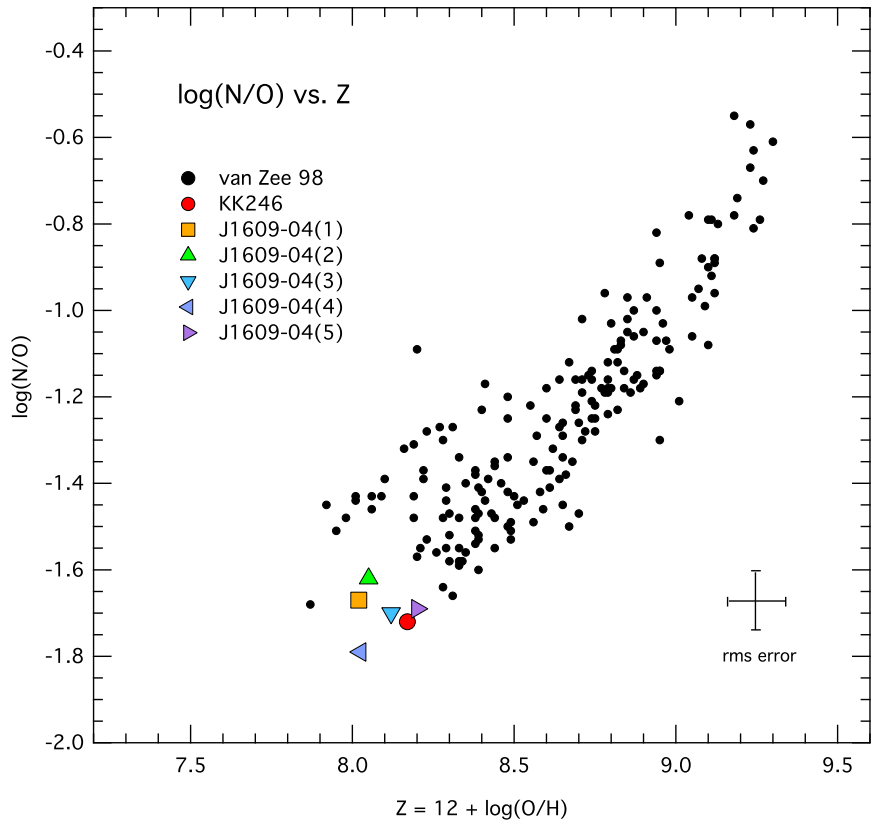}
\caption{log(N/O) vs. Oxygen metallicity for the objects observed here and data from \cite{Vanzee98}. Values of log(N/O) shown here correspond to an electron temperature of 11,250K.  The error bars indicate the rms error averaged over all observations.}\label{fig_9}
\end{figure}

In these data, as noted above, the value of Z (see table \ref{table_6}) was calculated using only the averaged values derived from diagnostics involving the ratio [NII]/[SII], as these diagnostics (at present) return the most accurate values ($\pm0.01$) of metallicity for test spectra using Mappings IV.

Figure 9 shows log(N/O) vs. Z, calculated as above, plotted with the extensive data from \cite{Vanzee98}. A linear fit to the distribution is usually taken as indicating objects with secondary nitrogen only (i.e., produced locally by previous generations of stars) \citep{Vanzee98,Vilacostas93}. The current observations smoothly extend to lower values the relationship between log(N/O) and Z in the van Zee data, and are consistent with there being mainly secondary nitrogen present in the \HII regions observed, and  low primary primary nitrogen, as defined by \cite{Vilacostas93}. There is no evidence of the transition ``knee'' from secondary (diagonal slope) to primary nitrogen (horizontal slope) at low (oxygen) metallicity.  This behavior appears reasonable for isolated dwarf galaxies with low or occasional star formation, and away from any possible enrichment of the intergalactic medium by large galaxies and galaxy mergers. 

The low (N/O) values are not fully consistent with the ``no winds'' models proposed by \cite{Gavilan13}, which strongly suggest a secondary to primary transition at metallicities found in the objects reported here.  This may indicate that the models need to take into account recent star formation, and  different initial mass functions (IMF) than previously considered.  It is conceivable that relatively low stellar masses in these \HII regions may be subject to ``small number statistics'', which make IMF modelling difficult. The previously reported scatter of log(N/O)  at low metallicity may be another feature of small number statistics, as proposed by \cite{Carigi08}.

\FloatBarrier

\subsection{Mass : metallicity}

Mass vs metallicity behaviour is one of the important evolutionary diagnostics for galaxies. It has been extensively mapped for larger galaxies \citep[e.g.,][]{Tremonti04}, but it is less well known for dwarf galaxies \citep{Lee06}.  Exploring it was one of the motivations for the SIGRID sample \citep{Nicholls11}. Figure \ref{fig_9} shows the stellar masses vs metallicities for KK246 (red) and J1609-04 (yellow, summed over all regions) plotted with data on similar low mass objects from \cite{Lee06} for comparison. Metallicity for the Lee objects was taken from several sources of good quality spectra, estimated using the $T_e$ method (and older atomic data).  The values for KK246 and J1609-04 have been reduced by 0.2 dex to compensate for the typical difference between strong line and $T_e$ method metallicities. Figure \ref{fig_9} shows that KK246 and J1609-04 follow the same trend as Lee et al.'s non-void dwarf objects, although, as has been found for void galaxies  \citep{Pustilnik11a}, J1609-04 falls a little below the trend.

\begin{figure}[htpb]
\centering
\includegraphics[width=0.6\hsize]{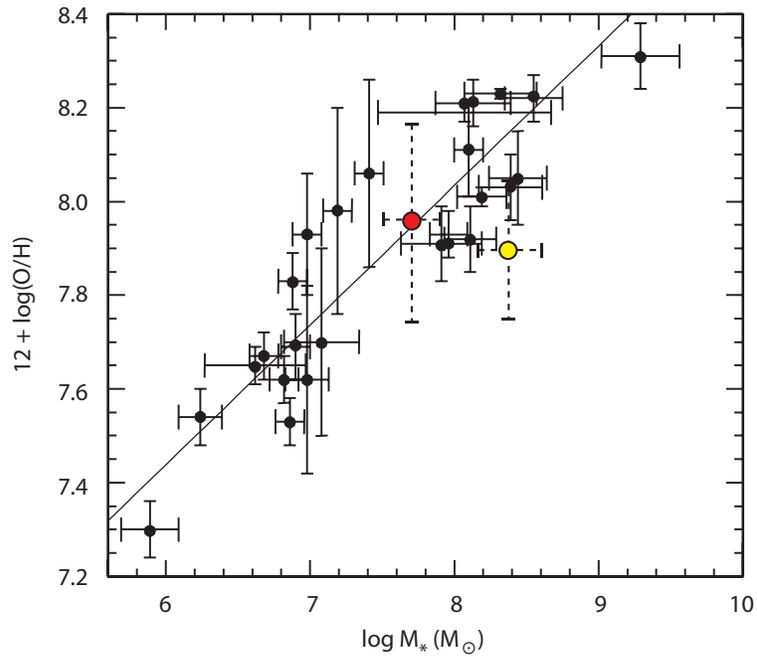}
\caption{Oxygen abundance vs. stellar mass for KK246 (red) and J1609-composite (yellow), plotted with $T_e$ method data for 25 nearby (D$<5$ Mpc) dwarf irregular galaxies from \cite{Lee06}. Our metallicity measurements are offset by -0.2 dex to compensate for the systematic difference between strong line and $T_e$ methods. The least-squares fit (solid line) applies only to the Lee et al. data.}\label{fig_9}
\end{figure}

\subsection{Mass vs light}

\begin{figure}[htpb]
\centering
\includegraphics[width=0.6\hsize]{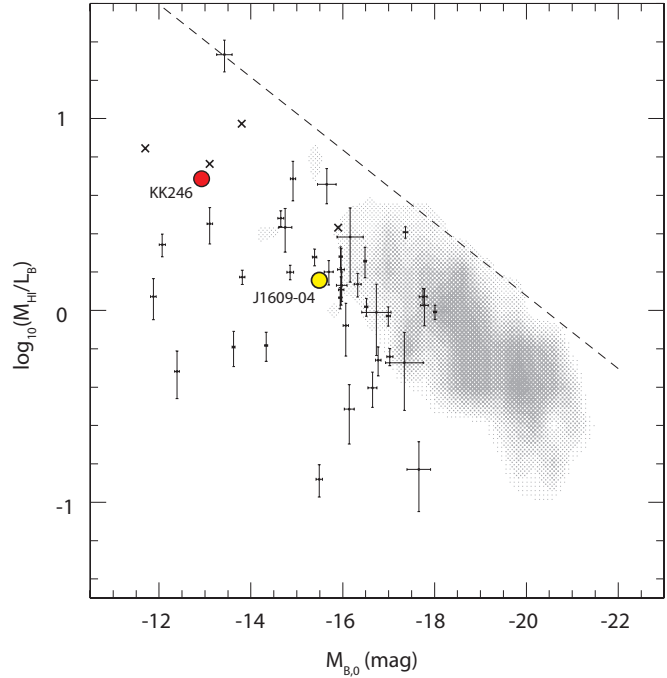}
\caption{Mass-to-light ratio vs absolute magnitude: log$_{10}$($\mathcal{M}_{HI}$/L$_B$) vs. M$_{B,0}$ for KK246 (red) and J1609-composite (yellow), plotted on data from Figure 2 from \cite{Warren07}. Error bars for KK246 and J1609-04 are within the plotted circles. The gray hatched area shows the HIPASS Bright Galaxy Gatalog objects, the objects with error bars are from Warren et al., and the crosses are other objects from Warren et al. from the literature.}\label{fig_10}
\end{figure}

Another question of interest is how the mass to light ratio vs. absolute magnitudes of these two galaxies compare with other similar galaxies. Are they intrinsically low luminosity objects? Values for  M$_{\rm H\textsc{i}}$/L$_B$ are shown in Tables 1 and 2: For KK246 the value is 4.66 and for J1609-04, 1.45.   In Figure \ref{fig_10} we plot KK246 and J1609-04 on data from Figure 2 from \cite{Warren07}. The shaded area shows 752 more luminous galaxies from the HIPASS Bright Galaxy Catalog. The dashed line shows the locus of an upper limit for the \HI mass-to-light ratio, as a function of luminosity.  The individual galaxies marked with error bars are those discussed by \cite{Warren07} and comprise late-type galaxies with a wide range of physical parameters. It is clear that there is nothing extreme about the two objects presented here---their M$_{\rm H\textsc{i}}$/L$_B$ ratios are typical of the 38 late-type galaxies with a range of mass-to-light ratios discussed by \cite{Warren07}.

\FloatBarrier

\subsection{Other studies of void galaxies}

The subject of galaxies in voids in the Local Volume has received increasing attention recently \citep{Pustilnik11a, Pustilnik11b,Kreckel12}.  \cite{Pustilnik11a, Pustilnik11b} have investigated galaxies in the Lynx-Cancer Void using their own and SDSS data, and find that in general, the metallicities are lower by $\sim$30\% than for higher density regions. In a study using SDSS data for galaxies in Local Volume voids, \cite{Weygaert11} and \cite{Kreckel12} found that, despite predictions from simulations, there was no significant population of \HI-rich low-luminosity galaxies filling the voids. They also found evidence of cold-mode accretion of \HI in some objects. Both studies included late-type spiral galaxies as well as dwarf irregulars.  For 48 Lynx-Cancer Void galaxies from \cite{Pustilnik11a}, the listed metallicity varies between 7.14 and 8.36.  These values are measured using the $T_e$ method, or using a semi-empirical method based on it, using older atomic data, and are likely underestimated by $\sim$0.2 dex compared to strong line metallicities \citep{Lopez12, Nicholls13}. 

\begin{figure}[htpb]
\centering
\includegraphics[width=0.6\hsize]{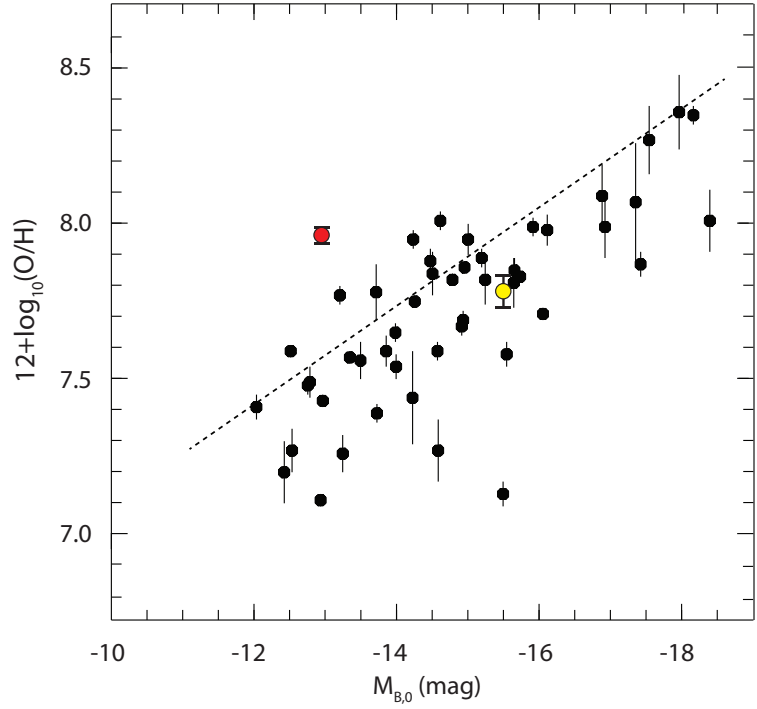}
\caption{Metallicity vs. M$_{B,0}$ for KK246 (red) and J1609-composite (yellow), plotted on data for Lynx-Cancer Void galaxies from \cite{Pustilnik11a}, offset vertically by -0.2 dex to allow for the $T_e$/strong line difference.}\label{fig_11}
\end{figure}

Figure \ref{fig_11} shows metallicity vs. M$_{B,0}$ for KK246 and J1609-04 plotted in the data for 42 Lynx-Cancer Void galaxies from \cite{Pustilnik11a}. The metallicities for KK246 and J1609-04 have been reduced by 0.2 dex to reflect the discrepancy between $T_e$ metallicities and strong line metallicities. The dashed line is for dI galaxies from \cite{Lee03}. \mbox{J1609-04} is typical of the other galaxies plotted, and lies near the trend defined by the dIrrs. KK246 is somewhat more metal rich, or conversely, less luminous in the B-band,  than the others.  This may be explained if the galaxy is nearing a ``post-starburst'' phase, with fewer blue stars present than would be expected from the current metallicity.

\FloatBarrier

\section{Discussion}

KK246 is the most isolated dwarf galaxy in the Local Volume, and J1609-04 is almost as isolated.  Thus their star formation tells us about the intrinsic behaviour of a gas-rich dwarf galaxy left to its own devices. It is apparent that these two objects represent quite normal dwarf irrregular galaxies. Their metallicities, nitrogen-to-oxygen ratios,  mass-to-light ratios and luminosities are typical for their masses. They appear to have evolved in the absence of external disturbance or enrichment in their local environments, so we may conclude that isolated dwarf galaxies do not need to be members of groups of clusters, or to have been influenced tidally to stimulate star formation and evolve normally, if they have a supply of \HI.  They build metallicities that are intrinsic and typical for their size. This suggests that, as a class, isolated dwarf irregular galaxies may not be good indicators of primordial abundances.

Although it has been suggested that small galaxies may not have sufficient mass to retain their supernova outflows, \cite{Bland-Hawthorn11} showed that full 3D modelling assuming a clumpy medium indicated that a Dark Matter halo (virial radius) mass of 3$\times 10^6$ M$_\odot$ is sufficient for a galaxy to gravitationally bind its supernova debris, and thus continue star formation while retaining chemical signatures laid down in the earlier epochs. Further evidence for this is that virtually no HIPASS \HI object is known without an accompanying galaxy or galaxy-component \citep{Koribalski04, Doyle05, Meurer06, Karachentsev08}. This was also demonstrated using disk evolutionary modelling by \cite{Taylor05}.   This implies that anywhere a concentration of neutral hydrogen occurs, a (dwarf) galaxy will evolve. It also indicates that the nebular metallicities of the two galaxies in this study reflect fully the star formation that has occurred in them.

From direct analysis of HST WFC3 photometry \citep{Karachentsev06}, KK246 shows more than one star forming episodes.  The old stellar population implied by H-band luminosity  suggests it has been forming stars over a protracted period. H-band imagery suggests that the same is true for J1609-04.    It is not clear how frequently star forming episodes occur in galaxies like these, and they may only happen occasionally.  From studies of isolated blue compact dwarf (BCD) galaxies, \cite{Sanchez08} have concluded that there is a duty cycle of 30:1 (quiescence to star burst). To judge from the BCD population in the SIGRID sample \citep{Nicholls11}, and the presence of at least two stellar populations in KK246 and J1609-04, this may be an underestimate. A better figure may be 10:1 or less, especially for less intense starburst episodes.

The individual  star-forming regions in J1609-04 vary in metallicity between  8.0 and 8.2. The metallicity of the regions do not correlate with their intrinsic luminosities, but this could be explained by the regions being of slightly different ages. The \HI region surrounding KK246 implies a plentiful supply of neutral hydrogen available for cold mode accretion. Such accretion has been observed in the Void Galaxy Survey in some galaxies \citep{Weygaert11}, and in Sag DIG in the Local Group \citep{Young97}.

From this, one may postulate a narrative for the development of small isolated galaxies. If they form early in the history of the Universe, with continuing  \HI available for cold inflow accretion, they will exhibit periodic star formation and evolve normally.  Their stellar populations will span a range of ages.  Supernova ejecta and stellar winds from the older stars will enrich the interstellar medium of the galaxy and a substantial fraction of the metallicity will be retained.  As a result the nebular metallicities will grow with time and will not be exceptionally low. While \cite{Pustilnik07} suggest that galaxies form later and evolve more slowly in regions of low density, this appears inconsistent with the old stellar populations in the two objects discussed here.

We also speculate that extremely low metallicity objects such as 1Zw18 \citep{Izotov99} and SBS 0335-052W \citep{Izotov05} are probably exceptional, and the absence of significant older stellar populations indicates that they have most likely formed recently from  pristine gas. The details of the older stellar population of 1Zw18 are not completely settled \citep{Garnett97,Aloisi99,Papaderos02}, but it has been claimed that there is no stellar population older than $\sim$500Myr  \citep{Hunt03,Izotov04}.  The question is still in dispute \citep{Annibali13, Papaderos12}. The irregular nature and complex kinematics of the \HI regions surrounding 1Zw18 \citep{Vanzee98} and to a lesser extent, SBS 0335-052W \citep{ Ekta09}, suggests that they are still in the process of forming. They may be the modern analogs of the so-called ``Faint Blue Galaxies'' at redshifts 1 $<$ z $<$ 2 \citep{Babul96,Ellis97}.  This has also been suggested by \cite{Papaderos12}. The ``bloated'' objects noted in the SIGRID sample \citep{Nicholls11} may be similar objects.  We have already observed some of these, and we plan to present data on their metallicities in a subsequent paper. 

%%%Check Janice Lee's 2009 SFR paper re H$\alpha$ and FUV SFR to see how J1609-04 and KK246 compare.

\section{Conclusions}

In this paper we present the first spectroscopic and nebular metallicity analysis of the two isolated dwarf gas-rich irregular galaxies, [KK98]246 and MCG-01-41-006. These objects were observed at resolutions of 3000 and 7000, with very good signal-to-noise ratios, using the WiFeS IFU spectrograph on the ANU 2.3 metre telescope at Siding Spring. To measure the metallicities, we have used the $T_e$ direct method (for the MCG galaxy) and the latest strong line methods from \cite{Dopita13}. We find that despite their location within or at the edge of the Local Void, their metallicities are no lower than other similar less isolated dwarf galaxies, with values of 8.17$\pm$0.01 and 8.04$\pm$0.05, respectively (and 7.96$\pm$0.06 for J1609-04 (2) and (5) using the $T_e$ method).  We suggest that isolated dwarf galaxies like these, with available inflows of \HI gas, will evolve normally, without the need for interactions with other galaxies. As a result, we believe that dwarf galaxies with exceptionally low metallicities are not the norm in voids, and that isolated dwarf irregular galaxies do not necessarily provide insights into the primordial intergalactic medium.

\begin{acknowledgments}
The authors wish to thank the referee for constructive comments, particularly relating to the nitrogen-to-oxygen abundance ratio. David Nicholls wishes to thank Prof.~Gary Da~Costa for useful discussions on stellar populations.  Mike Dopita acknowledges the support of the Australian Research Council (ARC) through Discovery  project DP0984657. This work was funded in part by the Deanship of Scientific Research (DSR), King Abdulaziz University, under grant No. (5-130/1433 HiCi). The authors acknowledge this financial support from KAU.
\end{acknowledgments}

\end{document}